\documentstyle[pra,aps]{revtex}
\begin{document}
\draft
\title{Quantum Dirac field without vacuum energy divergence}
\author{Ruo Peng WANG}
\address{
Physics Department, Peking University, Beijing 100871, P.R.China
}
\date{\today}

\maketitle
\begin{abstract}
A quantum Dirac field theory with no divergences of vacuum energy is presented. The vacuum energy divergence is eliminated by removing a extra degree of freedom of the Dirac fields. The conditions for removing the extra degree of freedom, expressed in the form of a conservation law and an orthogonality relation, define another spin $1/2$ field with the same rest mass that is just the antifermion field. The anticommutation relations for fermion and antifermion fields are imposed by this conservation law. Both fermion and antifermion fields have only states with positive energies due to the orthogonality relation. The expressions of the charge density and the current density are obtained. The charge conservation law is established. A form for canical quantization applicable to both Boson and Dirac fields is introduced.

\end{abstract}


The vacuum energy divergence is one of the divergences that occur in quantum
field theory. For the Dirac field this divergence is closely related to the
interpretation of negative-energy solutions of the Dirac equation. In
present quantum field theory, the annihilation operator of a fermion with 
negative energy is identified with the creation operator of an antifermion with
corresponding positive energy \cite{gmr}. In such a view, the vacuum state 
contains infinite number of fermions with negative energies, as firstly 
suggested by Dirac \cite{drc1}. A consequence of this assumption is that the 
vacuum  would have infinitely large negative energy and infinitely large 
negative charge \cite{gre}. Generally, this divergences are 
considered harmless, and are removed by redefining corresponding
operators with normal-ordered ones\cite{mandl}. 
However, in a mathematically well behaved theory, these divergences should 
not occur at all \cite{wu}. 
In this paper I introduce a quantum Dirac 
field theory, in which the total energy and the total charge of Dirac fields 
vanish in the vacuum state. 

The Dirac equation for particles of spin $1/2$ and rest mass $m$ is
\cite{drc2}

\begin{equation}\label{drc}
	\Bigl[i \hbar(\frac{\partial}	
	{\partial t} + c \vec {\alpha} \cdot \nabla) 
	-mc^2 \beta \Bigr] \psi_F(x)=0,
\end{equation}
in which $\psi_F(x)$ is a four components spinor, and the matrices $\vec {\alpha}$ and $\beta$ satisfy the following 
conditions:

\begin{equation}\label{com1}
	\left.
	\begin{array}{lll}
	\alpha^\dag_{r}=\alpha_{r}, &\alpha^2_{r}=1, 
	& [\alpha_{r},\alpha_{s}]_+=0 \; (r \neq s) \\
	\beta^\dag=\beta, &\beta^2=1, &	[\alpha_{r},\beta]_+=0
	\end{array} \right\}
	r,s=1,2,3,
\end{equation}
where the anticommutator is defined by

\begin{equation}
	[A,B]_+ \equiv AB+BA .
\end{equation}

The Dirac equation (\ref{drc}) allows both positive and negative energy solutions. Therefore the Dirac field defined only by Eq. (\ref{drc}) has one more degree of freedom than a real fermion field. To describe correctly real fermion fields, we must remove this extra degree of freedom. Removing of the extra degree of freedom is done by introducing a linear dependence between the components of the spinor field $\psi_F(x)$. This dependence must be covariant under Lorentz transformations, and it is convenient to write this relation in the following form:

\begin{equation}\label{rem}
	\frac{\partial}{\partial t}(\psi_A^T(x) \psi_F(x))
+\nabla \cdot (\psi_A^T(x) c \vec \alpha \psi_F(x)) =0,
\end{equation}
with $\psi_A(x)$ an arbitrary four components spinor. The convariance of the relation (\ref{rem}) requests $\psi_A(x)$ to transform like $\psi_F^*(x)$ under Lorentz transformations. In fact, the relation (\ref{rem}) and the Dirac equation (\ref{drc}) imply that $\psi_A(x)$ satisfies the same equation as $\psi_F^*(x)$:
 
\begin{equation}\label{drca} 
	\Bigl[i \hbar(\frac{\partial}{\partial t} +
	c \vec {\alpha}^T \cdot \nabla)  
	+mc^2 \beta^T \Bigr] \psi_A(x)=0,
\end{equation} 

It it easy to verify that the conditions (\ref{com1}) are also satisfied by
matrices $\vec {\alpha}^\prime$ and $\beta^\prime$ defined by

\begin{equation}
	\vec {\alpha}^\prime=\vec {\alpha}^T,\;
	\beta^\prime=-\beta^T.
\end{equation}
Therefore the spinor $\psi_A(x)$ describes a particle of spin $1/2$ and rest mass $m$ too. Usually, we refer to $\psi_F(x)$ as the fermion field and to $\psi_A(x)$ the antifermion field. 

There is a conservation law for the field $\psi_F(x)$:

\begin{equation}\label{cvs}
	\frac{\partial}{\partial t}(\psi_F^\dag (x) \psi_F(x))
+\nabla \cdot (\psi_F^\dag(x) c \vec \alpha \psi_F(x)) =0.
\end{equation}

Thus to ensure that the relation (\ref{rem}) being nontrivial, $\psi_A(x)$ and $\psi_F^*(x)$ must have no common component, that means the following orthogonality relation must be satisfied:

\begin{equation}\label{ort}
	\lim_{V \rightarrow \infty} 
	\frac{1}{V}\int_V \psi_A^T(x) \psi_F(x) d^3 \vec x=0.
\end{equation}

Eqs. (\ref{drc}) and (\ref{drca}) can be derived from the Lagrangian density 

\begin{eqnarray}\label{lgd}
	{\cal L}(x) &=& \psi_F^{\dag}(x)
	\Bigl[i \hbar(\frac{\partial}
	{\partial t} + c \vec {\alpha} \cdot \nabla) 
	-mc^2 \beta \Bigr]\psi_F(x)
	\nonumber \\
	&&+ \psi_A^{\dag}(x)
	\Bigl[i \hbar(\frac{\partial}
	{\partial t} + c \vec {\alpha}^T \cdot \nabla) 
	+mc^2 \beta^T \Bigr]\psi_A(x).
\end{eqnarray}

Relation (\ref{rem}) may be considered as a conservation law. To allow such a conservation law, the Lagrangian density (\ref{lgd}) must be invariant under certain transformation.
Let's consider the following transformation:

\begin{equation}\label{rot}
	\left.
	\begin{array}{l}
	\psi_F(x) \longrightarrow 
\psi_F^\prime(x)= \psi_F(x)+\epsilon_1 \psi_A^*(x)
	\\
	\psi_A(x) \longrightarrow \psi_A^\prime(x)= \psi_A(x) - 
	\epsilon_2 \psi_F^*(x)
	\end{array} \right\},
\end{equation}
where $\epsilon_1$ and $\epsilon_2$ are two infinitesimal parameters. According to the relation (\ref{ort}), we must have

\begin{equation}\label{orti}
	\lim_{V \rightarrow \infty} 
	\frac{1}{V} \Bigl[
	\epsilon_1 \int_V \psi_A^\dag(x) \psi_A(x) d^3 \vec x
	-\epsilon_2 \int_V \psi_F^\dag(x) \psi_F(x) d^3 \vec x
	\Bigr]=0.
\end{equation}

Under the transformation (\ref{rot}), the Lagrangian density (\ref{lgd}) is 
changed to

\begin{eqnarray}\label{lgdb}
	{\cal L}'(x) &=& {\cal L}(x) 
	+\epsilon_1 \psi_F^\dag(x)\Bigl[i \hbar(\frac{\partial}
	{\partial t} + c \vec {\alpha} \cdot \nabla) 
	-mc^2 \beta \Bigr]\psi_A^*(x)
	\nonumber \\
	&&+\epsilon_1^* \psi_A^T(x)\Bigl[i \hbar(\frac{\partial}
	{\partial t} + c \vec {\alpha} \cdot \nabla) 
	-mc^2 \beta \Bigr]\psi_F(x)
	\nonumber \\	
	&&-\epsilon_2 \psi_A^\dag(x)\Bigl[i \hbar(\frac{\partial}
	{\partial t} + c \vec {\alpha}^T \cdot \nabla) 
	+mc^2 \beta^T \Bigr]\psi_F^T(x)
	\nonumber \\
	&&-\epsilon_2^* \psi_F^T (x)\Bigl[i \hbar(\frac{\partial}
	{\partial t} + c \vec {\alpha}^T \cdot \nabla) 
	+mc^2 \beta^T \Bigr]\psi_A(x).
\end{eqnarray}
It would be possible to obtain the relation (\ref{rem}) from the Lagrangian density invariance condition, ${\cal L}(x) ={\cal L}^\prime(x)$, if we have $\epsilon_1=\pm \epsilon_2$. But the relation (\ref{orti}) is not satisfied in the case with $\epsilon_1 = - \epsilon_2$. Thus, only $\epsilon_1=\epsilon_2$ is allowed. To get the relation (\ref{rem}) under that circumstance, the following anticommutation relations must hold for Dirac fields:

\begin{equation}\label{com}
	\left.
	\begin{array}{c}
	\psi^\dag_p(x) O \psi_p(x) = -\psi^T_p(x) O^T \psi^{\dag T}_p(x) 
	\\
	\psi^T_a(x) O \psi_p(x) = -\psi^T_p(x) O^T \psi_a(x),\;(a \neq p)
	\end{array}
	\right\}a,p = F,A
\end{equation}
with $O=\vec \alpha, \beta$ or $1$. According to relations (\ref{lgdb}) and(\ref{com}), we have

\begin{eqnarray}\label{lgdc}
	{\cal L}'(x) &=& {\cal L}(x) 
	+\epsilon_1 i \hbar \Bigl[ \frac{\partial}
	{\partial t} \bigl(\psi_F^\dag(x)\psi_A^*(x) \bigr) + \nabla \cdot 	\bigl(\psi_F^\dag(x) 
	c \vec \alpha \psi_A^*(x) \bigr) \Bigr]
	\nonumber \\
	&&+\epsilon_1^* i \hbar \Bigl[ \frac{\partial}
	{\partial t} \bigl(\psi_A^T(x)\psi_F(x) \bigr) + \nabla \cdot 	\bigl(\psi_A^T(x) 
	c \vec \alpha \psi_F(x)\bigr) \Bigr].
\end{eqnarray}
Applying then the condition of Lagrangian density invariance in Eq. (\ref{lgdc}), the relation (\ref{rem}) is obtained.

We have already seen that the antifermion field $\psi_A(x)$ and the complex conjugate of the fermion field $\psi_F(x)$ satisfy the same equation. Thus a spinor function that satisfies Eq. (\ref{drca}) can be either a solution for the antifermion Dirac equation (\ref{drca}), or the complex conjugate of a solution for the fermion Dirac equation (\ref{drc}). It may not be the both at the same time due to the condition (\ref{ort}). Grouping solutions of the Dirac equation (\ref{drc}) or (\ref{drca}) into fermion and antifermion fields is rather arbitrary, and the only restriction is the condition (\ref{ort}). The arbitrariness in solution grouping should not affect the interaction between a fermion-antifermion fields pair and other fields. Therefore the charge and current densities of a fermion-antifermion fields pair must be independent on the grouping of solutions of the Dirac equation. By combining the conservation laws (\ref{rem}) and (\ref{cvs}), we may obtain the following expressions for the charge and current densities:

\begin{equation}\label{chrg}
	\left.
	\begin{array}{l}
	\rho_q(x)=q\{\psi_F^{\dag}(x)[\psi_F(x)+ \psi_A^*(x)]
	-[\psi_F^{T}(x) + \psi_A^{\dag}(x)]\psi_A(x)\}
	\\
	\vec j_q(x)=qc \{\psi_F^{\dag}(x) \vec \alpha 
	[\psi_F(x)+ \psi_A^*(x)] 
	-[\psi_A^{\dag}(x) + \psi_F^{T}(x)] 
	\vec \alpha^T \psi_A(x)\}
	 \end{array} \right\},
\end{equation}
where $q$ is the charge of the fermion. The conservation law for the charge 

\begin{equation}
	Q = \int \rho_q(x) d^3 \vec x,
\end{equation}
can be expressed in the form of the continuity equation

\begin{equation}
	\frac{\partial \rho_q}{\partial t}+\nabla \cdot \vec j_q=0.
\end{equation}

The Dirac fields $\psi_F(x),\psi_A(x)$ can be expanded into plan wave solutions. We have

\begin{equation}
	\psi_F(x)+ \psi^*_A(x)=\sum_{\vec k,s}\frac{1}{\sqrt{V}}
	e^{i[\vec{k} \cdot \vec{x}-\hbar^{-1}E(\vec k)t]}
	u_s{(\vec{k})} c_s(\vec k) + \sum_{\vec k,s}\frac{1}{\sqrt{V}}
	e^{-i[\vec{k} \cdot \vec{x}-\hbar^{-1}E(\vec k)t]}
	v^*_s{(\vec{k})} d^\dag_s(\vec k),
\end{equation}
where $s=1,2$ is the label for spin states, and
$u_s{(\vec{k})}$ and $v_s{(\vec{k})}$ satisfy the following equations 

\begin{equation}\label{drc_1}
	\Bigl[E(\vec k)
	- c\hbar \vec k \cdot \vec {\alpha} 
	-m c^2 \beta \Bigr]u_s(\vec k)=0
\end{equation}

\begin{equation}\label{drca_1}
	\Bigl[E(\vec k)
	- c\hbar \vec k \cdot \vec {\alpha}^T 
	+m c^2 \beta^T \Bigr]v_s(\vec k)=0
\end{equation}
and orthonormality relations

\begin{equation}
	u_s^\dag (\vec{k})u_{s^\prime}(\vec{k}) = \delta_{ss^\prime},\;
	v_s^\dag (\vec{k})v_{s^\prime}(\vec{k}) = \delta_{ss^\prime},\;
	v_s^T (\vec{k})u_{s^\prime}(-\vec{k}) =0,
\end{equation}
with $ E(\vec k)=\sqrt{\hbar^2 |\vec k|^2+m^2c^4}$. We have a physical choice for solution grouping:

\begin{equation}\label{expd}
	\psi_F(x)=\sum_{\vec k,s}\frac{1}{\sqrt{V}}
	e^{i[\vec{k} \cdot \vec{x}-\hbar^{-1}E(\vec k)t]}
	u_s{(\vec{k})} c_s(\vec k),	\;\;
	\psi_A(x)=\sum_{\vec k,s}\frac{1}{\sqrt{V}}
	e^{i[\vec{k} \cdot \vec{x}-\hbar^{-1}E(\vec k)t]}
	v_s{(\vec{k})} d_s(\vec k).
\end{equation}
One may observe that under such a choice, both of $\psi_F(x),\psi_A(x)$ contain only positive-energy solutions. We have another way for expressing this solution grouping:

\begin{eqnarray}
	\psi_F(x)=\sum_{\vec k,s}\frac{1}{V} \int_V d^3 \vec x^\prime
	u_s{(\vec{k})} u^\dag_s{(\vec{k})} 
	e^{i\vec{k} \cdot (\vec{x}-\vec x^\prime)}
	( \psi_F(x)+ \psi^*_A(x))
	\\
	\psi_A(x)=\sum_{\vec k,s}\frac{1}{V} \int_V d^3 \vec x^\prime
	v_s{(\vec{k})} v^\dag_s{(\vec{k})} 
	e^{i\vec{k} \cdot (\vec{x}-\vec x^\prime)}
	( \psi_A(x)+ \psi^*_F(x))
\end{eqnarray}
For non-quantized Dirac
field, the relations (\ref{com}) become

\begin{equation}\label{com2}
	[c_s (\vec{k}), c^\dag_{s^\prime}(\vec{k}^\prime)]_+
	= [d_s (\vec{k}), d^\dag_{s^\prime}(\vec{k}^\prime)]_+
	=0,
\end{equation}
and

\begin{equation}\label{com3}
	[c_s^\dag (\vec{k}), d^\dag_{s^\prime}(\vec{k}^\prime)]_+
	= [c_s (\vec{k}), d_{s^\prime}(\vec{k}^\prime)]_+
	=0.
\end{equation}
To obtain the Hamiltonian density, we need to calculate the conjugate fields 
of $\psi_F(x),\psi^\dag_F(x),\psi_A(x),\psi^\dag_A(x)$. We have

\begin{equation}\label{dag}
	\left.
	\begin{array}{l}
	\pi_F(x)=\frac{\partial{\cal L}(x)}{\partial \dot{\psi}_F(x)}
	=i\hbar \psi_F^\dag(x) ,\;
	\tilde{\pi}_F(x)=\frac{\partial{\cal L}(x)}
	{\partial \dot{\psi}_F^\dag(x)}=0
	\\
	\pi_A(x)=\frac{\partial{\cal L}(x)}{\partial \dot{\psi}_A(x)}=
	i\hbar \psi_A^\dag(x)  ,\;
	\tilde\pi_A(x)=\frac{\partial{\cal L}(x)}
	{\partial \dot{\psi}_A^\dag(x)}=0
	\end{array} \right\}.
\end{equation}
Then, the Hamiltonian density is given by

\begin{eqnarray}
	{\cal H}(x) &=& \psi_F^{\dag}(x)
	\bigl(-i \hbar c \vec {\alpha} \cdot \nabla 
	+mc^2 \beta \bigr)\psi_F(x)
	\nonumber \\
	&&+ \psi_A^{\dag}(x)
	\bigl(-i \hbar c \vec {\alpha}^T \cdot \nabla 
	-mc^2 \beta^T \bigr)\psi_A(x). 
\end{eqnarray}
Due to the anticommutation property of Dirac fields, the relations (\ref{dag}) are not obvious, and we need an explicit definition of the derivative of Lagrangian.  The following definition is used:

\begin{equation}\label{der}
	\frac{\partial{\cal L}(x)}{\partial \dot{\psi}_F(x)}
	\equiv \lim_{\Delta \dot{\psi}_F(x) \rightarrow 0}
	\left( -\frac{1}{\Delta \dot{\psi}_F(x)}\Delta {\cal L}(x) \right)
	=\lim_{\Delta \dot{\psi}_F(x) \rightarrow 0}
	\left(\Delta {\cal L}(x) \frac{1}{\Delta \dot{\psi}_F(x)} \right).
\end{equation}

The canonical quantization condition $ [q, p] = i \hbar $ can be written as

\begin{equation}\label{quan}
	\frac{\partial}{\partial \dot{q}}[q, L] = i \hbar,
\end{equation}
where $L$ is the Lagrangian of the system. By using the expressions (\ref{expd}), the Lagrangian of the system can be calculated. We have
\begin{equation}
	L=\sum_{\vec k,s}\left( i\hbar c^\dag_s(\vec k, t) \dot {c}_s(\vec k, t)
	+ i\hbar d^\dag_s(\vec k, t) \dot {d}_s(\vec k, t)
	-E(\vec k) c^\dag_s(\vec k, t)c_s(\vec k, t) 
	-E(\vec k) c^\dag_s(\vec k, t)c_s(\vec k, t) \right),
\end{equation}
with

\begin{equation}
	c_s(\vec k, t)= c_s(\vec k)e^{-i\hbar^{-1}E(\vec k)t},\;\;
	d_s(\vec k, t)= d_s(\vec k)e^{-i\hbar^{-1}E(\vec k)t}.
\end{equation}
According to the quantization condition (\ref{quan}), we have

\begin{equation}
	\lim_{\Delta \dot{c}_s(\vec k,t) \rightarrow 0}
	\left(-\frac{1}{\Delta \dot{c}_s(\vec k,t)} \bigl(c_s(\vec k, t) 	c^\dag_s(\vec k, t)\Delta \dot {c}_s(\vec k,t)
	- c^\dag_s(\vec k, t)\Delta \dot {c}_s(\vec k,t) c_s(\vec k, t)
  	\bigr) \right)=1.
\end{equation}
From the relations (\ref{com2}), we have
\begin{equation}\label{comd}
	[\Delta \dot {c}_s (\vec{k}), c^\dag_{s^\prime}(\vec{k}^\prime)]_+
	=0,
\end{equation}
thus

\begin{equation}
	c_s(\vec k, t) c^\dag_s(\vec k, t)\Delta \dot {c}_s(\vec k,t)
	- c^\dag_s(\vec k, t)\Delta \dot {c}_s(\vec k,t) c_s(\vec k, t)
	=-\Delta \dot {c}_s(\vec k,t)\bigl(c_s(\vec k, t) c^\dag_s(\vec k, t)+ 
	c^\dag_s(\vec k, t) c_s(\vec k, t)\bigr),
\end{equation}
and the quantization condition for Dirac fields becomes

\begin{equation}
	[c_s(\vec k, t), c^\dag_s(\vec k, t)]_+=1.
\end{equation}
Similarly,

\begin{equation}
	[d_s(\vec k, t), d^\dag_s(\vec k, t)]_+=1.
\end{equation}
The above relations are equivalent to

\begin{equation}\label{comq}
	[c_s (\vec{k}), c^\dag_{s^\prime}(\vec{k}^\prime)]_+
	= [d_s (\vec{k}), d^\dag_{s^\prime}(\vec{k}^\prime)]_+
	= \delta_{ss^\prime}\delta_{\vec k \vec k^\prime}.
\end{equation}
One may observe the fact that the relation (\ref{comd}) is consistent with the quantization conditions.

The quantization relations are obtained by using the first definition for the derivatives of Lagrangian. But the two definitions in (\ref{der}) are equivalent, therefore we must obtain the same quantization relations basing on the second definition. Indeed, the same quantization relations are derived if the following anticommutation relations hold:

\begin{equation}\label{com4}
	[ c_s (\vec{k}), c_{s^\prime}(\vec{k}^\prime) ]_+ =
	[ c^\dag_s (\vec{k}), c^\dag_{s^\prime}(\vec{k}^\prime) ]_+ =
	[ d_s (\vec{k}), d_{s^\prime}(\vec{k}^\prime) ]_+ =
	[ d^\dag_s (\vec{k}), d^\dag_{s^\prime}(\vec{k}^\prime) ]_+ =0 .
\end{equation}
The quantization conditions (\ref{comq}) are invariant under the transformation (\ref{rot}). To keep the invariance of anticommutation relations (\ref{com4}) under the same transformation, we must have

\begin{equation}
	[ c_s (\vec{k}), d^\dag_{s^\prime}(\vec{k}^\prime) ]_+ =
	[ c^\dag_s (\vec{k}), d_{s^\prime}(\vec{k}^\prime) ]_+ =0.
\end{equation}

The coefficients $c_s(\vec k), d_s(\vec k)$ can be interpreted the as operators for annihilating a fermion, respectively, an antifermion in the state characterized by a wave vector $\vec k$ and spin label $s$, and the coefficients $c^\dag _s(\vec k), d^\dag _s(\vec k)$ can be considered as operators for creating a fermion,respectively, an antifermion in a given state.

By using plan wave expansion, one obtains the following expression for the 
Hamiltonian operator:

\begin{eqnarray}
	H &=& \int {\cal H}(x) d^3 \vec x
	\nonumber \\
	&=&\sum_{\vec{k}s}E(\vec{k}) \bigl[N_{Fs}(\vec{k})
	+N_{As}(\vec{k}) \bigr].
\end{eqnarray}

The vacuum state $|0 \rangle$ is defined by \cite{mandl}

\begin{equation}
	c_s (\vec{k}) |0 \rangle = d_{s}(\vec{k}) |0 \rangle =0,\;
	\mbox{for all} \;\vec k, \;\mbox{and} \; s=1,2.
\end{equation}
It is obviously that the numbers of fermion and antifermion in the vacuum 
state are zero:

\begin{equation}
	N_{Fs} (\vec{k}) |0 \rangle = N_{As}(\vec{k}) |0 \rangle =0,\;
	\mbox{for any} \;\vec k, \;\mbox{and} \; s,
\end{equation}
and consequently, the total energy of the quantum Dirac field in the vacuum 
state is null.

The total charge of the quantum Dirac field can also be calculated. We find

\begin{equation}
	Q = \sum_{\vec{k}s}q \bigl[N_{Fs}(\vec{k})
	-N_{As}(\vec{k}) \bigr].
\end{equation}
It is evident that the total charge of the quantum Dirac field also vanishes 
in the vacuum state. 

In summary, I have introduced a quantum Dirac field theory, in which the 
vacuum energy divergence is eliminated by removing an extra degree of freedom of the Dirac fields. The conditions for removing the extra degree of freedom of the Dirac field define another spin $1/2$ field with the same rest mass that is refered as the antifermion field. These conditions can be expressed in the form of a conservation law and an orthogonality relation. The conservation law requests the invariance of the Lagrangian density under certain transformation, and the anticommutation relations for fermion and antifermion fields are imposed by this invariance. Free fermion and antifermion fields have only states with positive energies, because the existence of negative energy states would be in conflict with the orthogonality relation. Therefore the vacuum energy divergence does not occur. The expressions for charge and current density are obtained, and the charge conservation law for fermion and antifermion fields is established as well. A form for canical quantization applicable to both Boson and Dirac fields is also introduced.

\end{document}